\documentclass[showpacs,preprintnumbers,amsmath,amssymb,aps,prd,nofootinbib,cite,eqsecnum,11pt]{revtex4}
\usepackage{amsmath}
\usepackage[dvipdfm]{hyperref}

\makeatletter
\renewcommand{\p@subsection}{}
\makeatother

\makeatletter
\newcommand{\xequal}[2][]{\ext@arrow 0055{\equalfill@}{#1}{#2}}
\def\equalfill@{\arrowfill@\Relbar\Relbar\Relbar}
\makeatother

\renewcommand{\thesection}{\arabic{section}}
\renewcommand{\thesubsection}{\arabic{section}.\arabic{subsection}}

\makeatletter
\renewcommand{\theequation}{\arabic{section}.\arabic{equation}}
\@addtoreset{equation}{section}
\makeatother

\newcommand{\chushi}[1]{ }

\newcommand{\angleN}[1]{ \langle #1 \rangle }

\newcommand{\roundLR}[1]{ \left( #1 \right) }
\newcommand{\roundN}[1]{ ( #1 ) }

\newcommand{\roundB}[1]{ \biggl( #1 \biggr) }

\newcommand{\squareLR}[1]{ \left[ #1 \right] }

\newcommand{\absLR}[1]{ \left| #1 \right| }
\newcommand{\absN}[1]{ | #1 | }

\let\calccommentout\iffalse 
\let\calcshow\iftrue 

\newcommand{\eq}[1]{\begin{equation}\begin{split} #1 \end{split}\end{equation}}

\newcommand{\eqna}[1]{\begin{eqnarray} #1 \end{eqnarray}}

\newcommand {\mathsym}[1]{{}}
\newcommand {\unicode}[1]{{}}

\begin{document}

\title{Phenomenological Construction of New Dictionaries\\for Holographic Conductors }

\author{Hironori Hoshino}
\affiliation{
Department of Physics, Chuo University,
Tokyo 112-8551, Japan
}
\author{Shin Nakamura}
\affiliation{
Department of Physics, Chuo University,
Tokyo 112-8551, Japan
}


\begin{abstract}
We propose new dictionaries for holographic conductors that enable us to compute carrier densities and mean velocities of charge carriers in the system. The carrier density, which differs from the charge density, is the total number density of both the positive and the negative charge carriers. The mean velocity is the mean value of the velocities of all charge carriers. These quantities are not conjugate to the sources that are given by boundary values of bulk fields, and we cannot compute them by using the conventional method in holography.
In the present work, we introduce a phenomenological model of charge transport, and we establish the dictionary by comparing the results of the phenomenological model and those from the holography. We show that the mean velocity agrees with the velocity of an analog black hole on the worldvolume of the probe D-brane, and it can be read from the spectrum of the fluctuations.
\end{abstract}
\maketitle

\section{introduction}
\label{introduction}

Investigation of nonequilibrium phenomena beyond the linear response regime is important in many aspects. For example, nonlinear charge transport is a key property of electric devices. From the theoretical point of view, better understanding of such nonequilibrium phenomena may provide us a useful insight into construction of nonequilibrium statistical mechanics. 
 
Recently, the gauge/gravity correspondence \cite{Maldacena:1997re,Gubser:1998bc}, which is also referred to as the AdS/CFT correspondence or holography, has been paid attention in its application to nonequilibrium physics \cite{Hubeny:2010ry}. The gauge/gravity correspondence is a map between a microscopic theory of gauge particles and a gravity theory on a curved spacetime. Although the gauge-theory side is defined at the level of the microscopic theory, the macroscopic physics naturally appears just by solving the classical equations of motion in the gravity side. For example, thermodynamics of the gauge particles is realized as thermodynamics of the black hole geometry that is obtained by solving the Einstein's equation. This suggests that the process of coarse graining that connects the microscopic theory and the macroscopic physics is somehow encoded in the classical theory of gravity. This remarkable nature of gravity makes us ambitious to apply the gauge/gravity correspondence for the systems out of equilibrium in the nonlinear regime. 

Indeed, a method to compute nonlinear electric conductivity of gauge-particle systems has been proposed within the framework of the gauge/gravity correspondence \cite{Karch:2007pd}.
Interesting nonlinear properties, such as negative differential conductivity (NDC) \cite{Nakamura:2010zd}, has been obtained by using this method. Novel nonequilibrium phase transitions associated with the nonlinear conductivity have also been discovered within this framework \cite{Nakamura:2012ae}. 
In addition to the nonlinear characteristics of the gauge-particle systems,
the gauge/gravity correspondence tells us basic properties of nonequilibrium steady states (NESS). One interesting feature of the gravity dual is that basic properties of NESS can be described by analog black holes on the probe D-branes (or probe fundamental strings) embedded into the dual geometry. 
For example, an effective temperature of NESS is realized as the Hawking temperature of the analog black hole \cite{Sonner:2012if, Nakamura:2013yqa, Hoshino:2014nfa}\footnote{Effective temperatures for Langevin systems have been analyzed in terms of the analog black holes in \cite{Gursoy:2010aa}.}.
In this sense, once we establish a precise map between the parameters of the analog black holes and the physical quantities of the corresponding gauge-particle systems, we may reach better understanding of NESS in terms of the analog black holes.

In Ref.~\cite{Nakamura:2010zd}, it has been pointed out that the pair creation process of the positively charged particles (positive carriers) and the negatively charged particles (negative carriers) is essential for the realization of the NDC. 
However, we still lack better understanding of the mechanism for the NDC.
We expect that the physical quantities such as the {\it number density of charge carriers} (which we call {\it carrier density} in this paper), that is defined as the number density of the positive carriers {\it plus} that of the negative carriers, may provide us useful information on the mechanism.
This is because the carrier density is determined by the balance between the pair creation and the annihilation of the charge carriers.
However, to the best knowledge of the authors, any method to compute the carrier density has not been known within the framework of the gauge/gravity correspondence.
The reason is that the carrier density is not a conserved quantity and not conjugate to any sources that are given by boundary values of bulk fields.
Hence, we cannot apply the GKP-Witten prescription \cite{Gubser:1998bc} for computation of carrier densities.


In the present paper, we provide new dictionaries to compute physical quantities that are not directly obtained from the GKP-Witten prescription.
We introduce a phenomenological model of the underlying microscopic process, and we compare the prediction of the model with the results of the gauge/gravity correspondence. 
This enables us to find the relationship between the physical quantities introduced in the model and those computable from the GKP-Witten prescription.
For example, we obtain dictionaries to compute the carrier densities, friction coefficients, and the mean velocities of the charge carriers.



We get a further benefit from the foregoing framework.
It has been found that analog black holes that describe NESS at finite current densities have a ``velocity'' \cite{Kim:2011qh}.
The velocity is read at the horizon of the analog black hole following the standard black hole physics. However, as far as the authors understand, there is still room for discussion on the physical interpretation of this velocity in terms of the corresponding gauge-particle systems.
Our framework gives us a natural interpretation of the velocity of the analog black hole as the mean velocity of the charge carriers.
We show how the mean velocity can be read from the spectrum of the fluctuations in the corresponding NESS, as well.

The organization of the present paper is as follows.
In Section \ref{sec:classicalpicture_1}, we introduce a phenomenological model for the charge transport.
In Section \ref{sec:drudelikewithholography}, we compute the current density by using the gauge/gravity correspondence.
In Section \ref{sec:dictionary}, we compare the results of Section \ref{sec:classicalpicture_1} and Section \ref{sec:drudelikewithholography}, and propose the new dictionaries.
We analyze the spectrum of the fluctuations in Section \ref{sec:fluct}.
We conclude in Section \ref{sec:discuss}.

\section{Phenomenological picture}
\label{sec:classicalpicture_1}

\subsection{Phenomenological Model}

In this section, we introduce a phenomenological model that describes charge transport in our system. 
We consider a system in which a constant external electric field $\vec{E}$ and a constant external magnetic field $\vec{B}$ are acting on charge carriers whose charge density is $\rho$. The charge carriers interact with neutral particles that form a heat bath of temperature $T$. We assume that the charge carriers have either charge $+1$ or $-1$ and they have a common mass $m$ irrespective of their charge. 
The pairs of the positive charge carrier and the negative charge carrier can recombine, or can be created by the external electric field. This means that the number density $n$ of the charge carriers is determined as a function of the control parameters $(\rho, \vec{E}, \vec{B}, T)$ by nontrivial dynamics of the system. In other words, $n$ reflects the effects of the recombination and the pair creation of the charge carriers.
We define the number density of the positive charge carriers to be $\rho_{+}$, and that of the negative charge carriers to be $\rho_{-}$, so that we have
\begin{eqnarray}
\rho&=&\rho_{+}-\rho_{-},
\label{rho-def}\\
n&=&\rho_{+}+\rho_{-}.
\label{n-def}
\end{eqnarray}
We refer $n$ as {\it carrier density} in this paper, whereas $\rho$ as charge density.

We assume that the $i$-th charge carrier obeys a Langevin equation
\begin{eqnarray}
\frac{d \vec{p}_{i}}{dt}=-\hat{\gamma}_{i} \vec{v}_{i}+\vec{F}(\vec{v_{i}})+\xi_{i},
\label{Langevin}
\end{eqnarray}
where $\vec{p}_{i}$ and $\vec{v}_{i}$ are the momentum and the velocity of the $i$-th charge carrier, respectively. $\hat{\gamma}_{i}$ is the friction coefficient associated with the $i$-th charge carrier. $\vec{F}(\vec{v_{i}})$ and $\xi_{i}$ are the external force and the random force acting on the $i$-th charge carrier, respectively. The external force we consider is a constant electromagnetic force acting on the charge:
\begin{eqnarray}
\vec{F}(\vec{v}_{i})=q_i(\vec{E}+\vec{v}_{i}\times \vec{B}),
\label{Lorentz}
\end{eqnarray}
where $q_i=\pm 1$ denotes the charge of the charge carrier under consideration.

In the conventional model, $\hat{\gamma}_{i}$ is set to be a constant that is independent of $\vec{v}_{i}$. However, we relax this assumption so that $\hat{\gamma}_{i}$ can depend on $|\vec{v}_{i}|$. Furthermore, $\hat{\gamma}_{i}$ may also depend on the velocities of other charge carriers through the interaction. This means that the Langevin equation couples with that of other charge carriers.

Let us consider NESS.
We assume that the time scale we consider is long enough compared with the microscopic time scale so that we can average the microscopic fluctuation in time to define a macroscopic quantity. In NESS, the macroscopic quantity after averaging the fluctuation does not evolve in time.
Then $\langle \frac{d}{dt}\vec{p}_{i} \rangle=\langle \xi_{i} \rangle=0$,
and we obtain
\begin{eqnarray}
\langle \hat{\gamma}_{i}\vec{v}_{i} \rangle=q_i(\vec{E}+\langle \vec{v}_{i}\rangle\times \vec{B}),
\label{mean-Langevin}
\end{eqnarray}
where $\langle {\cal O} \rangle$ denotes the time average of quantity $\cal O$. 
We assume that the averaged values are common to identical particles, and the equations (\ref{mean-Langevin}) are boiled down to two equations,
\begin{eqnarray}
\langle \hat{\gamma}_{+}\vec{v}_{+} \rangle&=&(\vec{E}+\langle \vec{v}_{+}\rangle\times \vec{B}),
\label{p-eq}\\
\langle \hat{\gamma}_{-}\vec{v}_{-} \rangle&=&-(\vec{E}+\langle \vec{v}_{-}\rangle\times \vec{B}),
\label{m-eq}
\end{eqnarray}
where those with $+$ ($-$) subscript denotes the quantities associated with the positive (negative) charge carriers.

We assume that NESS is determined when we specify the control parameters. This means that $\langle \hat{\gamma}_{\pm}\vec{v}_{\pm} \rangle$ and $\langle \vec{v}_{\pm} \rangle$ are given as functions of the control parameters $(\rho, \vec{E}, \vec{B}, T)$. 
Let us rearrange the left-hand sides of (\ref{p-eq}) and (\ref{m-eq}) in such a way that
\begin{eqnarray}
\langle \hat{\gamma}_{\pm}\vec{v}_{\pm} \rangle
=\gamma_{\pm} \langle \vec{v}_{\pm} \rangle,
\label{meanfield-2}
\end{eqnarray}
where we have extracted $\langle \vec{v}_{\pm} \rangle$ and we have introduced new functions $\gamma_{\pm}$ of $(\rho, \vec{E}, \vec{B}, T)$.
For notational simplicity, we omit $\langle \ \rangle$ from now on. Then, (\ref{p-eq}) and (\ref{m-eq}) are given as 
\begin{eqnarray}
\gamma_{\pm}\vec{v}_{\pm} =\pm(\vec{E}+\vec{v}_{\pm}\times \vec{B}).
\label{balance}
\end{eqnarray}
These equations express the balance between the friction force and the external force.

With the foregoing notations, the current density $\vec{J}=(J^{x}, J^{y}, J^{z})$ is given by
\begin{eqnarray}
\vec{J}=\rho_{+}\vec{v}_{+}-\rho_{-}\vec{v}_{-},
\label{J-def}
\end{eqnarray}
and the mean velocity $\vec{v}_{\mbox{\scriptsize m}}$ of the charge carriers is {defined} by
\eq{
\vec{v}_{\mbox{\scriptsize m}}
=
	\frac{  \rho_+ \vec{v}_+ + \rho_- \vec{v}_- }{n}	
.
\label{eq:averagevelocityintheclassicalpicture}
}
Note that our definition of the mean velocity is different from $\vec{J}_{n_{q}}/\rho$ which is referred to as the ``net velocity'' in \cite{Kim:2011qh}.
Here, $\vec{J}_{n_{q}}$ is the contribution to the current due to the existence of the doped charge carriers dragged by the external electric field.

\subsection{Condition for $\gamma_+=\gamma_-$}
\label{sec:classicalpicture_generalresult}

Solving (\ref{balance}) with respect to $\vec{v}_{\pm}$, we obtain
\begin{eqnarray}
	\vec{v}_{\pm}=\pm\frac{(\vec{E}\cdot\vec{B})\vec{B}+\gamma_{\pm}\left(\gamma_{\pm}\vec{E}\pm \vec{E}\times \vec{B}\right)}{\gamma_{\pm}(B^{2}+\gamma_{\pm}^{2})},
\label{eq:vpm}
\end{eqnarray}
where $B^{2}=\vec{B}\cdot\vec{B}$.
Substituting this into $\vec J$, we reach the following relationship
\eqna{
	&&(\vec{J}\cdot\vec{B})\vec{J}\cdot(\vec{E}\times \vec{B})+\rho(\vec{J}\times \vec{E})\cdot(\vec{E}\times \vec{B})
	\nonumber \\
	&&=\left(\frac{1}{B^{2}+\gamma_{+}^{2}}-\frac{1}{B^{2}+\gamma_{-}^{2}}\right)
	\left[\frac{1}{\gamma_{+}}+\frac{1}{\gamma_{-}}\right]\rho_{+}\rho_{-}(\vec{E}\cdot\vec{B})(\vec{E}\times \vec{B})^{2}.
\label{pre_sym_condition}
}

Let us consider the case where
\eq{
	(\vec{E}\cdot\vec{B})(\vec{E}\times \vec{B})^{2}\neq 0
.
\label{eq:conditionforEandB}
}
Then (\ref{pre_sym_condition}) means that 
$\gamma_{+}=\gamma_{-}$ or $\rho_{+}\rho_{-} = 0$, if 
\begin{eqnarray}
	(\vec{J}\cdot\vec{B})\vec{J}\cdot(\vec{E}\times \vec{B})+\rho(\vec{J}\times \vec{E})\cdot(\vec{E}\times \vec{B})=0.
\label{sym_condition}
\end{eqnarray}
When (\ref{sym_condition}) holds, we obtain 
\eqna{
	{\gamma}^2
&=&
	\rho
	\frac{
	\roundN{	\vec{E}\times \vec{B} }^2
	}{
	\vec J
	\cdot
	\roundN{	\vec{E}\times \vec{B} }
	}
	-
	B^2
\label{eq:gammaunderthecondition1}
,
\\
	n
&=&
	\frac{
	{\gamma}
	\vec{J}\cdot \vec{B}
	}{
	{ \vec{E} \cdot \vec{B} }
	}
=
	\frac{
	\vec J
	\cdot 
	\vec E
	}{
	{\gamma}
	v^2
	}
,
\label{eq:dtilapp1}
}
where we have renamed $\gamma_\pm$ and $v_{\pm}^{2} (= \angleN{v_+}^2= \angleN{v_-}^2)$ to $\gamma$ and $v^2$, respectively\footnote{The derivation of (\ref{pre_sym_condition}), (\ref{eq:gammaunderthecondition1}), and (\ref{eq:dtilapp1}) is given in Appendix \ref{sec:derivationofgpeqgm}.}.
We can check that $n$ is finite even at the limit of $v^{2}=0$.
The second equality in (\ref{eq:dtilapp1}) indicates that the power consumed in the system per unit time per unit volume is given by the product of the energy loss of a single carrier per unit time and the carrier density $n$.  
The point is that, if the condition (\ref{sym_condition}) is satisfied, the positive charge carriers and the negative charge carriers share the common friction {coefficient} $\gamma$ and the common absolute value of the velocity $|v|$.


For later use, let us exhibit expressions for the case where
$\vec E = (E_x,0,0)$ and $\vec B = (B_x,0,B_z)$.
We employ this field configuration without loss of generality.
This setting satisfies (\ref{eq:conditionforEandB}).
The condition (\ref{sym_condition}) is written as
\eq{
	B_x J^x J^y + B_z J^y J^z + \rho E_x J^z 
=
	0
.
\label{eq:conditionforgammapeqgammam}
}
Under this condition, the components of $\vec{v}_{+}$ and $\vec{v}_{-}$ are related as $v_{+}^x=-v_{-}^x \equiv v^{x}$, $v_{+}^y=v_{-}^y \equiv v^{y}$, $v_{+}^z=-v_{-}^z \equiv v^{z}$.
Hence the current density (\ref{J-def}) becomes
\eq{
	\vec J
&=
	\begin{pmatrix}
	n {v^x}
	\\
	\rho {v^y}
	\\
	n {v^z}
	\\
	\end{pmatrix}
.
\label{eq:Jbyv}
}
We also find that (\ref{eq:vpm}), (\ref{eq:gammaunderthecondition1}) and (\ref{eq:dtilapp1}) yield
\eqna{
	v^{2}
&=&
	\frac{E_x^2 \left(B_x^2+{\gamma}^2\right)}{{\gamma}^2
	\left(B^2+{\gamma}^2\right)}
\label{eq:vsquare}
,
\\
	{\gamma}^2
&=&
	-
	\rho
	\frac{
	E_x B_z 
	}{
	J^y
	}
	-
	B^2
,
\label{eq:gammagene}
\\
	n
&=&
	\frac{
	J^x
	E_x
	}{
	\gamma
	v^2
	}
.
\label{eq:dtilgene}
}
The mean velocity (\ref{eq:averagevelocityintheclassicalpicture}) is given by
\eq{
	\vec{v}_{\mbox{\scriptsize m}}
&=
	\frac{ 1 }{n}	
	\begin{pmatrix}
	\rho
	{v^x}
	\\
	n
	{v^y}
	\\
	\rho
	{v^z}
	\\
	\end{pmatrix}
.
\label{eq:definitionofaveragev}
}
The density dependence in (\ref{eq:definitionofaveragev}) is understood as follows.
Along the directions of $x$ and $z$, the positive carriers and the negative carriers are driven in the opposite directions with the same speed.
Hence $v_m^x$ and ${v_m^z}$ are proportional to $\rho=\rho_{+}-\rho_{-}$.
In the $y$ direction, the positive carriers and the negative carriers equally contribute to ${v_m^y}$, hence the factor ${1}/{n}$ is canceled.

\section{Non-linear conductivity from gauge/gravity correspondence}
\label{sec:drudelikewithholography}

We have studied the phenomenological model so far.
In order to compare the results from the phenomenological model and predictions of holography, we compute nonlinear conductivities in holographic models. It will be found that (\ref{eq:conditionforgammapeqgammam}) holds in general in a wide range of holographic models.

\subsection{Gravity dual}
\label{sec:setupofholography}
Our gravity dual is constructed on a 10-dimensional space-time.
In the radial direction, which we call $u$, there is a boundary located at $u=u_{\text{b}}$.\footnote{Typically, $u_{\text{b}}=0$ for the Fefferman-Graham type coordinates, and $u_{\text{b}}=\infty$ for the Schwarzschild type coordinates.}
The boundary is a $(d+1)$-dimensional Minkowski space-time with $d \geq 3$, the coordinates of which are  $(t,\vec x) = (t,x^{1},x^{2},x^{3},\cdots,x^{d})$ where we denote $(x^{1},x^{2},x^{3})=(x,y,z)$.
We assume that the remaining $(8-d)$-dimensional space is compact.
To describe a finite temperature system, we consider a black hole geometry dual to the heat bath.
The horizon is located at $u=u_H$, and the Hawking temperature is identified with the temperature of the heat bath. We assume that the metric is given by
\eq{
	ds^2
&=
	g_{tt}(u)
	dt^2
	+
	g_{xx}(u)
	d \vec{x}^2
	+
	g_{uu}(u)
	du^2
	+
	d \Omega^2
,
\label{eq:metricofgeneralpqnwithLbyz}
}
where $d\Omega^2$ gives the line-element along the compact manifold.
The metric (for the non-compact part) depends only on $u$, and they are regular between the horizon and the boundary.
{In our convention, $g_{tt} < 0$ outside the horizon.}
At the boundary, $-g_{tt}$, $g_{xx}$ and $g_{uu}$ positively diverge. $g_{tt}$ vanishes at the horizon whereas $g_{xx}$ remains finite there.

The sector of the charge carriers is described by a probe D($q+1+n$)-brane\footnote{The formalism presented here is a generalization of that in \cite{Ammon:2009jt}.}, where $d \geq q \geq 3$.
We employ the probe approximation.
We assume that the $(q+1)$-dimensional spatial directions of the worldvolume extend in the $u$-direction and the directions of $x^{1}, x^{2}, x^{3}, \cdots, x^{q}$.
The $n$-dimensional part wraps an $n$-dimensional subspace of the $(8-d)$-dimensional compact manifold, hence $8-d \geq n$.
The action for the probe D($q+1+n$)-brane is given by
\eq{
	S_{D(q+1+n)}
&=
-
	T_{D(q+1+n)}
	\int d^{q+1+n} \zeta
	e^{-\phi}
	\sqrt{-\det
	\roundLR{
	h_{ab}
	+
	2\pi \alpha'
	{F}_{ab}
	}
	}
,
\label{eq:DBIaction}
}
where $T_{D(q+1+n)}$ is the tension of the brane, and $\zeta^a$ are the worldvolume coordinates.
We assume that the dilaton field $\phi$ depends only on $u$.
The constant $\alpha'$ is related to the t'Hooft coupling $\lambda$\footnote{
When the bulk geometry is the D$p$-brane background, $\lambda = 2N_c g_s (2\pi)^{p-2}  \alpha'^{(p-3)/2} $ \cite{Itzhaki:1998dd}.
}.
The induced metric $h_{ab}$ and the U(1) gauge field strength ${F}_{ab}$ on the brane are defined by
\eq{
	h_{ab}
&=
	\partial_a X^\mu
	\partial_b X^\nu
	g_{\mu \nu}
,
\qquad
	{F}_{ab}
=
	\partial_a {A}_{b}
	-
	\partial_b {A}_{a}
,
}
where $X^\mu(\zeta)$ represents the configuration of the D($q+1+n$)-brane, and $A_a$ is the U(1) gauge field.
We employ the static gauge.
In this paper, we consider the cases where the Wess-Zumino term is not switched on.

We consider homogeneous NESS where the macroscopic quantities do not depend on the space-time coordinates on the boundary: we assume that $F_{ab}$ and $X^\mu$ depend only on $u$.
For the gauge field, we employ the following ansatz:
\eq{
	A_x(t,u) &= -E_xt +h_x(u)
,
\qquad
	A_y(x,u) = B_z x +h_y(u)
,
\qquad
	A_z(y,u)
=
	B_x y
	+
	h_z(u)
,
\label{eq:ansatzes}
}
and the other components of the gauge field are switched off in our gauge choice.

Under the present setup, $h_{ab}+(2\pi \alpha') F_{ab}$ is block diagonal
so that the action is factorized as
\eq{
	S_{D(q+1+n)}
&=
	-
	\int d t d \vec{x} d u
	\mathcal{L}
,
\qquad
	\mathcal{L}
=
	V
	\sqrt{
	-
	\squareLR{
	h
	+
	(2\pi\alpha')^2
	C_1
	+
	(2\pi\alpha')^4
	C_2
	}
	}
\label{eq:probeDqp1pnaction}
,
}
where $h=h_{tt}h_{xx}^{3}h_{uu}$, 
\eq{
	V
&=
	T_{D(q+1+n)}
	\int d^n \zeta
	e^{-\phi}
	h_{xx}^{(q-3)/2}
	\sqrt{h_{\Omega}}
,
\\
	C_1
&=
	h_{xx}^2 
	\squareLR{
	h_{xx}
	{A}_t'^2 
	+
	h_{tt}
	\roundLR{
	{A}_x'^2
	+
	{A}_y'^2 
	+
	{A}_z'^2 
	}
	}
	+
	h_{xx}
	h_{uu}
	\squareLR{
	h_{xx}
	{E}_x^2 
	+
	h_{tt}
	\roundLR{
	{B}_x^2
	+
	{B}_z^2
	}
	}
,
\\
	C_2
&=
	h_{xx}
	\left({B}_x^2+{B}_z^2\right)
	{A}_t'^2 
	+
	h_{tt} 
	\roundLR{
	{B}_x
	{A}_x' 
	+
	{B}_z
	{A}_z'
	}^2
	+
	h_{xx}
	{E}_x^2 
	\left({A}_y'^2+{A}_z'^2\right)
	+
	h_{uu}
	{E}_x^2 
	{B}_x^2
	-
	2 
	h_{xx}
	{E}_x {B}_z
	{A}_t' 
	{A}_y' 
,
}
and the prime denotes $\partial/\partial u$.
$\sqrt{h_{\Omega}}$ is the volume element of the worldvolume along the compact directions, which may depend on the brane configuration in the compact manifold. This means that $\sqrt{h_{\Omega}}$, hence $V$ contains dynamical degrees of freedom, as well as $h_{uu}$ does, and they are determined by the Euler-Lagrange equations.

Let us consider the dynamics of the gauge fields. 
The GKP-Witten prescription gives the following identifications~\cite{Karch:2007pd,O'Bannon:2007in,Ammon:2009jt}:
\eq{
	\frac{\delta \mathcal{L}}{\delta {A}_t'}
&=
	\rho
,
\qquad
	\frac{\delta \mathcal{L}}{\delta {A}_i'}
=
	J^i
\qquad
	(i=x,y,z)
,
\label{eq:integratedEOM}
}
where $(\rho,J^i)$ are the expectation values of the charge density and the current density, respectively.
These equations yield
\eq{
	{A}_t'^2
&=
	\frac{
	\absLR{h_{tt}}
	h_{zz} 
	}{
	\left(
	h_{xx}^2
	+
	(2\pi\alpha')^2
	{B}_x^2
	\right)^2
	}
	\frac{
	(
	\rho
	w
	-
	{B}_z
	a_1
	)^2
	}{
	w
	f
	-
	\frac{
	a_1^2
	}{
	h_{xx}^2
	+
	(2\pi\alpha')^2
	{B}_x^2
	}
	+
	\frac{
	a_2^2
	}{
	\absLR{h_{tt}}
	h_{xx} 
	-
	(2\pi\alpha')^2
	{E}_x^2
	}
	}
,
\\
	{A}_x'^2
&=
	\frac{
	h_{zz} 
	}{
	\absLR{h_{tt}}
	h_{xx}^2 
	}
	\frac{
	(
	{J^x}
	{w}
	-
	{B}_x 
	a_2
	)^2
	}{
	w
	f
	-
	\frac{
	a_1^2
	}{
	h_{xx}^2
	+
	(2\pi\alpha')^2
	{B}_x^2
	}
	+
	\frac{
	a_2^2
	}{
	\absLR{h_{tt}}
	h_{xx} 
	-
	(2\pi\alpha')^2
	{E}_x^2
	}
	}
,
\\
	{A}_y'^2
&=
	\frac{
	h_{zz} 
	}{
	\absLR{h_{tt}}
	h_{xx}^2 
	}
	\frac{
	(
	{J^y}
	{w}
	+
	{E}_x {a_1}
	)^2
	}{
	w
	f
	-
	\frac{
	a_1^2
	}{
	h_{xx}^2
	+
	(2\pi\alpha')^2
	{B}_x^2
	}
	+
	\frac{
	a_2^2
	}{
	\absLR{h_{tt}}
	h_{xx} 
	-
	(2\pi\alpha')^2
	{E}_x^2
	}
	}
,
\\
	{A}_z'^2
&=
	\frac{
	\absLR{h_{tt}}
	h_{zz} 
	}{
	\left(
	\absLR{h_{tt}}
	h_{xx} 
	-
	(2\pi\alpha')^2
	{E}_x^2
	\right)^2
	}
	\frac{
	(
	{J^z}
	{w}
	-
	{B}_z 
	{a_2}
	)^2
	}{
	w
	f
	-
	\frac{
	a_1^2
	}{
	h_{xx}^2
	+
	(2\pi\alpha')^2
	{B}_x^2
	}
	+
	\frac{
	a_2^2
	}{
	\absLR{h_{tt}}
	h_{xx} 
	-
	(2\pi\alpha')^2
	{E}_x^2
	}
	}
,
\label{eq:squareofsolutinofgaugefields}
}
where ${w}$, ${f}$, ${a_1}$ and ${a_2}$ are defined as
\eq{
	w
	(u)
&=
	\absLR{h_{tt}}
	h_{xx}^3
	-
	(2\pi \alpha')^2
	h_{xx}^2
	{E}_x^2
	+
	(2\pi \alpha')^2
	\absLR{h_{tt}}
	h_{xx} 
	{B}^2
	-
	(2\pi \alpha')^4
	{E}_x^2
	{B}_x^2
,
\\
	f
	(u)
&=
	(2\pi\alpha')^4
	\absLR{h_{tt}}
	h_{xx}^2
	V^2
	+
	\frac{
	\absLR{h_{tt}}
	h_{xx} 
	}{
	h_{xx}^2
	+
	(2\pi\alpha')^2
	{B}_x^2
	}
	{\rho}^2
	-
	{J^x}^2
	-
	{J^y}^2
	-
	\frac{
	\absLR{h_{tt}}
	h_{xx} 
	}{
	\absLR{h_{tt}}
	h_{xx} 
	-
	(2\pi\alpha')^2
	{E}_x^2
	}
	{J^z}^2
,
\\
	a_1
	(u)
&=
	(2\pi\alpha')^2
	\squareLR{
	\absLR{h_{tt}}
	h_{xx} 
	{B}_z 
	{\rho}
	+
	\left(
	h_{xx}^2
	+
	(2\pi\alpha')^2
	{B}_x^2
	\right)
	{E}_x
	{J^y}
	}
,
\\
	a_2
	(u)
&=
	(2\pi\alpha')^2
	\squareLR{
	\left(
	\absLR{h_{tt}}
	h_{xx}
	-
	(2\pi\alpha')^2
	{E}_x^2
	\right)
	{B}_x 
	{J^x} 
	+
	\absLR{h_{tt}}
	h_{xx} 
	{B}_z 
	{J^z}
	}
.
\label{eq:definitionofKs}
}
There is a point $u=u_*$ where $w(u_*)=0$, since $w$ is positive at the boundary whereas $w$ is negative at the horizon.
The condition $w(u_*)=0$ implies 
\eq{
	\absLR{h_{tt}}
	h_{xx}
	-
	(2\pi\alpha')^2
	{E}_x^2
	\Big|_{u=u_*}
&=
	-
	\frac{
	(2\pi\alpha')^2
	\absLR{h_{tt}}
	h_{xx} 
	{B}_z^2
	}{
	h_{xx}^2
	+
	(2\pi\alpha')^2
	{B}_x^2
	}
	\Bigg|_{u=u_*}
\leq
	0
,
\label{eq:negative}
}
since $h_{tt}\leq 0$ and $h_{xx}> 0 $.

For the reality of the gauge fields in (\ref{eq:squareofsolutinofgaugefields}), we need
\eq{
	{w}
	{f} 
	-
	\frac{
	{a_1}^2
	}{
	h_{xx}^2
	+
	(2\pi\alpha')^2
	{B}_x^2
	}
	+
	\frac{
	{a_2}^2
	}{
	\absLR{h_{tt}}
	h_{xx} 
	-
	(2\pi\alpha')^2
	{E}_x^2
	}
&\geq
	0
,
\label{eq:conditionforKchixia1}
}
for all region of $u$.
Especially, when $u=u_*$, (\ref{eq:conditionforKchixia1}) is satisfied if and only if
\eq{
	a_1
	(u_*)
&=
	{a_2}
	(u_*)
=
	0
,
\label{eq:asarezeroatustar}
}
where we have used (\ref{eq:negative}).
This means that the left-hand side of (\ref{eq:conditionforKchixia1}) is minimized at $u=u_*$, and hence we obtain
\eq{
	\partial_u
	\squareLR{
	{w}
	{f} 
	-
	\frac{
	{a_1}^2
	}{
	h_{xx}^2
	+
	(2\pi\alpha')^2
	{B}_x^2
	}
	+
	\frac{
	{a_2}^2
	}{
	\absLR{h_{tt}}
	h_{xx} 
	-
	(2\pi\alpha')^2
	{E}_x^2
	}
	}
	\Bigg|_{u=u_*}
=
	0
.
\label{eq:dercond}
}
Here, $\partial_u w (u_*)$, $\partial_u a_1 (u_*)$, $\partial_u a_2 (u_*)$, and $	
	\partial_u
	\squareLR{
	\absLR{h_{tt}}
	h_{xx} 
	-
	(2\pi\alpha')^2
	{E}_x^2
	}^{-1}
	|_{u=u_*}
$ are finite\footnote{Although the denominator of the third term in (\ref{eq:dercond}) vanishes at $B=0$, one finds that the third term is still finite by using (\ref{eq:definitionofKs}) and (\ref{eq:asarezeroatustar}).}, under our assumption. 
In addition, by using $w(u_*)=0$, (\ref{eq:asarezeroatustar}), and  (\ref{eq:dercond}), we find
\eq{
	{f} (u_*)
&=
	0
.
\label{eq:chiiszero}
}
(\ref{eq:asarezeroatustar}) and (\ref{eq:chiiszero}) give the current density as
\eq{
	J^i
&=
	\sigma^{ix}
	E_x
,
\label{eq:JinD3D7withB3}
}
where the conductivity tensor is given by
\eq{
	\sigma^{xx}
&=
	\frac{(2\pi \alpha')}{{h}_{xx}}
	\frac{
	{h}_{xx}^2+(2\pi \alpha')^2{B}_x^2 
	}{
	{h}_{xx}^2+(2\pi \alpha')^2{B}^2
	}
	\sqrt{
	{\rho}^2
	+
	(2\pi\alpha')^2
	{h}_{xx}
	\roundLR{
	{h}_{xx}^2
	+
	(2\pi \alpha')^2
	{B}^2
	}
	V^2
	}
	\Bigg|_{u=u_*}
,
\\
	\sigma^{yx}
&=
	-
	\frac{
	(2\pi \alpha')^2{B}_z {\rho} 
	}{
	{h}_{xx}^2
	+
	(2\pi \alpha')^2{B}^2
	}
	\Bigg|_{u=u_*}
,
\\
	\sigma^{zx}
&=
	\frac{
	(2\pi \alpha')^2{B}_x {B}_z 
	}{
	{h}_{xx}^2
	+
	(2\pi \alpha')^2{B}_x^2
	}
	\sigma^{xx}
	\Bigg|_{u=u_*}
.
\label{eq:relabetJ2andDandJ3andJ1}
}
Note that $V$ remains in $\sigma_{xx}$ at the limit of $\rho \to 0$.
This implies that $V$ contains contributions of pair creations as is discussed in \cite{Karch:2007pd}.

\section{New dictionary}
\label{sec:dictionary}
{Now we are ready to make a matching between the results of the model in Sec.~\ref{sec:classicalpicture_1} and those from holography in Sec.~\ref{sec:drudelikewithholography} to find new dictionaries.}
\subsection{General background}
\label{sec:result1}
Note that (\ref{eq:JinD3D7withB3}) with (\ref{eq:relabetJ2andDandJ3andJ1}) satisfies (\ref{eq:conditionforgammapeqgammam}).
Hence $\gamma_+=\gamma_-$ holds for the holographic models we have considered, and the discussions in Sec.~\ref{sec:classicalpicture_generalresult} apply.
We obtain
\eqna{
	{\gamma}
&=&
	\frac{{h}_{xx}(u_*)}{(2\pi \alpha')}
,
\label{eq:gammagene2}
\\
	v
&=&
	\absLR{\vec v}
=
	\sqrt{
	\frac{\absLR{h_{tt}(u_*)}}{h_{xx}(u_*)}
	}
,
\label{eq:sokudo2gene}
\\
	n
&=&
	2\pi\alpha'
	\frac{
	{J}^x
	{E}_x
	}{
	\absN{ {h}_{tt}(u_*) }
	}
=
	\sqrt{
	{\rho}^2
	+
	h_{xx} 
	\roundLR{
	h_{xx}^2
	+
	(2\pi \alpha')^2
	{B}^2
	}
	(2\pi \alpha')^2
	V^2
	}
	\Big|_{u=u_*}
.
\label{eq:dtilgene2}
}
We have also
\begin{eqnarray}
	\Gamma
&\equiv&
 	\gamma
 	\sqrt{1-v^2}
=
	\frac{1}{(2\pi\alpha')}
	\sqrt{
	{h}_{xx}(u_*)
	\roundB{
	{h}_{xx}(u_*)
	-
	\absLR{h_{tt}(u_*)}
	}
	}
~,
\label{eq:Gamma}
\\
	\rho_{+}\rho_{-}
&=&
	\frac{1}{4}
	\left[
	h_{xx} 
	\roundLR{
	h_{xx}^2
	+
	(2\pi \alpha')^2
	{B}^2
	}
	(2\pi \alpha')^2
	V^2
	\right]
	\Big|_{u=u_*}
,
\label{eq:dtilgene2-2}
\end{eqnarray}
that are equivalent to (\ref{eq:gammagene2}) and (\ref{eq:dtilgene2}), respectively.
{$\Gamma$ is a friction coefficient divided by the Lorentz factor.}
These are our new dictionaries.
One finds that $n$ approaches $\rho$ at the limit of $\rho/V (u_*) \to \infty$. This indicates that $n$ is equivalent to $\rho$ when the charge density effect is dominant over that of the pair creation.

Note that $w(u)$, hence $u_{*}$ as well, does not depend on $\rho$, $V$ and $h_{uu}$. Then we find the following remarkable facts: $\gamma$, $v$ and $\Gamma $ are independent of $\rho$ and the D-brane's configuration in the compact space.
They do not depend on the properties of the probe brane, and 
they are not sensitive to the properties of the sector of the charge carriers (such as the mass $m$ of the carriers) in the NESS at least under the probe approximation.\footnote{A related discussion is given in \cite{O'Bannon:2007in}.}  

For later use, we exhibit the $x$ component of the mean velocity (\ref{eq:definitionofaveragev}) at $B=0$:
\eq{
	v_m^x
&=
	\frac{ \rho}{n}	
	{v^x}
=
	\frac{\rho J^x }{
	(2\pi \alpha')^2 
	h_{xx}^{3}
	V^2
	+
	{\rho^2}
	}
	\Bigg|_{u=u_*}
,
\label{eq:vmandrhoj}
}
where we have used (\ref{eq:Jbyv}) and (\ref{eq:dtilgene2}) in the last equality. This will be identified with the velocity of the analog black hole on the worldvolume theory of the probe D-brane in Sec.~\ref{sec:OSMandvm}. This identification, together with (\ref{eq:vmandrhoj}), provides a dictionary for the mean velocity.

\subsection{Example: the D$p$-brane background}
\label{sec:result2}

Let us consider the case where the bulk geometry is given by the D$p$-brane background.
The D$p$-brane geometry is given by \cite{Itzhaki:1998dd} as

\begin{eqnarray}
	ds^2
=
	u^{\frac{7-p}{2}}
	\left[
	-\left(
	1-u_{H}^{7-p}/u^{7-p}
	\right)dt^2
	+d\vec{x}^{2}
	\right]
	+
	\frac{du^{2}}{
	u^{\frac{7-p}{2}}
	\left(
	1-u_{H}^{7-p}/u^{7-p}
	\right)}
	+
	u^{\frac{p-3}{2}}
	d\Omega^{2}_{8-p}
,
\label{eq:metricofgeneralpqnwithL}
\end{eqnarray}
where $d\Omega^{2}_{8-p}$ gives the metric on the unit $S^{8-p}$, and the dilaton $\phi$ is given by
\eq{
	e^{\phi}
&=
	e^{\phi_{0}} u^{(p-3)(7-p)/4}
,
}
with a constant $\phi_{0}$. The Hawking temperature is given by
\eq{
	T
&= 
	\frac{7-p}{4\pi}\: u_{H}^{\frac{5-p}{2}}
.
\label{eq:relabetzhandT}
}

In the present setup, we have 
\eq{
	h_{xx}
	(u_*)
&=
	u_{H}^{\frac{7-p}{2}}
	\sqrt{K}
,
\qquad
	h_{tt}
	(u_*)
=
	u_{H}^{\frac{7-p}{2}}\:
	\frac{
	1
	-
	K
	}{
	\sqrt{K}
	}
,
\label{eq:relabetweenhtth11andGEB}
}
where
\eq{
	K
&=
	\frac{1}{2}
	\squareLR{
	e_x^2
	-
	\roundLR{
	b_x^2
	+
	b_z^2
	}
	+
	1
	+
	\sqrt{
	\left(
	e_x^2
	-
	\roundLR{
	b_x^2
	+
	b_z^2
	}
	+
	1
	\right)^2
	+
	4
	\roundLR{
	b_x^2
	+
	b_z^2
	}
	+
	4 b_x^2
	e_x^2
	}
	}
.
}
The dimensionless quantities $e_x$ and $b_i$ are defined by
\eq{
	e_1
&=
	2 \pi \alpha'  {E}_1
u_{H}^{-\frac{7-p}{2}}
,
\qquad
	b_i
=
	2 \pi \alpha'  {B}_i
u_{H}^{-\frac{7-p}{2}}
.
\label{eq:dimlesseandbinDp}
} 
Note that the $K$ depends only on $\vec E$, $\vec B$, $T$ and $\lambda$, and hence $u_*$ does in the D$p$-D($q+1+n$) model.

{Substituting (\ref{eq:relabetweenhtth11andGEB}) into (\ref{eq:dtilgene2}) and (\ref{eq:Gamma}), we obtain }
\eqna{
	n
&=&
	\sqrt{
	{\rho}^2
	+
	\sqrt{K}
	\roundLR{
	K
	+
	{b}^2
	}
	(2\pi \alpha')^2
	u_{H}^{{7-p}}
	V^2(u_*)
	}~
,
\label{eq:DtildeinDp}
\\
	\Gamma
&=&
	\frac{u_{H}^{(7-p)/2}}{2\pi \alpha'}	
,
\label{eq:GammainDp}
}
where $b^2 = b_x^2+b_z^2$.
{The absolute value of the velocity (\ref{eq:sokudo2gene}) becomes }
\eq{
	v
&=
	\sqrt{
	\frac{\absLR{h_{tt}(u_*)}}{h_{xx}(u_*)}
	}
=
	\sqrt{
	\frac{
	K
	-
	1
	}{
	K
	}
	}
.
\label{eq:relabetvandFEB}
}
{We find $0 \leq v \leq 1$ since $0 \leq\absLR{h_{tt}}/h_{xx} \leq 1$: $v$ does not exceed the speed of light. }

We find a remarkable nature of the results in the models with the D$p$-brane background. In more general models in Sec.~\ref{sec:setupofholography}, $\Gamma$ may depend on $u_{*}$ hence on the external fields. However, in the D$p$-brane background, $\Gamma$ is independent of $u_{*}$ and it depends only on the heat bath temperature $T$ and the coupling constant: 
we share the common $\Gamma$ irrespective of $m$.
This means that our $\Gamma$ (\ref{eq:GammainDp}) is the same as that of the models for drag force~\cite{Herzog:2006gh,Gubser:2006bz,CasalderreySolana:2006rq}, where the model has only a single infinitely massive carrier.


\section{Fluctuation spectrum}
\label{sec:fluct}
In this section, we are going to study how the mean velocity
appears in the open-string metric on the probe D-brane, and how it contributes to the distribution of fluctuations in NESS. We consider only $\vec B=0$ cases, in this section.


\subsection{Open-string metric and mean velocity}
\label{sec:OSMandvm}
Let us look into the open string metric $G_{ab}=h_{ab}-(2\pi\alpha^{\prime})^{2}(Fh^{-1}F)_{ab}$ on the probe D-brane at $\vec B=0$.
The open string metric is diagonal in the absence of the field strength, since $h_{ab}$ is diagonal in our setup.
However, we have off-diagonal components $G_{ut}=G_{tu} \propto F_{tx} h^{xx} F_{xu}$ in the presence of $E_{x}$ and $J^{x}$. If the charge density $\rho$ is further turned on, $G_{xt}=G_{tx} \propto F_{tu} h^{uu} F_{ux}$ appears, because both $F_{ut} \propto \rho$ and $F_{ux} \propto J^x$ are non-vanishing through (\ref{eq:squareofsolutinofgaugefields}).

We consider only the $3\times 3$ part of the open-string metric on the $(t,x,u)$ coordinates.
We attempt to diagonalize ${G}_{ab}$ in the vicinity of $u_*$ by using the following coordinate transformation: 
\eq{
	\begin{pmatrix}
	d\hat{t}\\
	d \hat{x} \\
	d\hat{u} \\
	\end{pmatrix}
=
	\left(
	\begin{array}{c}
	dt
	+
	Q^{\hat{t}}_{~u}
	du 
	\\
	dx 
	+ 
	Q^{\hat{x}}_{~t}
	dt  
	+ 
	Q^{\hat{x}}_{~u}
	du 
	\\
	du
	\end{array}
	\right)
,
\label{eq:diagonalizedminicoordinatecomponents}
}
where $Q^{\hat a}_{~b}$ are defined by
\eq{
	Q^{\hat{t}}_{~u}
&=
	\frac{
	{G}_{tx} {G}_{xu}-{G}_{xx}
	   {G}_{tu}
	}{
	{G}_{xt}^2-{G}_{xx} {G}_{tt}
	}
,
\qquad
	Q^{\hat{x}}_{~t}
=
	\frac{{G}_{xt}}{{G}_{xx}}
	\Bigg|_{u=u_*}
,
\qquad
	Q^{\hat{x}}_{~u}
=
	\frac{{G}_{xu}}{{G}_{xx}}
.
\label{eq:defofQs}
}
The metric $\mathcal{G}_{ab}=[(Q^{-1})^{T}GQ^{-1}]_{ab}$ on the new coordinates is given by
\eq{
	\mathcal{G}_{\hat{t} \hat{t}}
&=
	\frac{{G}_{tt}{G}_{xx}-{G}_{xt}^2}{{G}_{xx}}
	+O\left((u-u_{*})^{2}\right)
,
\qquad
	\mathcal{G}_{\hat{x}\hat{x}}
=
	{G}_{xx}
,
\qquad
	\mathcal{G}_{\hat{u} \hat{u}}
=
	\frac{
	-\det {G}
	}{
	{g}_{xx}^2
	\roundLR{
	{G}_{tt}
	{G}_{xx}
	-
	{G}_{xt}^2
	}
	}
	+O(1)
,
\label{eq:diagmetric}
}
where $\det G$ is the determinant of $G_{ab}$. ${G}_{tt}{G}_{xx}-{G}_{xt}^2$ is at the order of $u-u_{*}$, and vanishes at $u=u_{*}$. 
Other off-diagonal components are $O(u-u_{*})$. 

Note that $Q^{\hat{x}}_{~t}$ is defined in (\ref{eq:defofQs}) as a constant given by ${G}_{xt}/{G}_{xx}$ at $u=u_*$ so that we keep $\partial_t \partial_u \hat x = \partial_u \partial_t \hat x $. This induces the off-diagonal components at $u \not = u_*$.
The inverse of the metric is given by
\eq{
	\mathcal{G}^{\hat{t} \hat{t}}
&=
	\frac{{G}_{xx}}{{G}_{tt}{G}_{xx}-{G}_{xt}^2}
,
\qquad
	\mathcal{G}^{\hat{x}\hat{x}}
=
	\frac{1}{{G}_{xx}}
	+O[(u-u_{*})]
,
\qquad
	\mathcal{G}^{\hat{u} \hat{u}}
=
	\frac{{G}_{tt}{G}_{xx}-{G}_{xt}^2}{\det {G}}
,
\\
	\mathcal{G}^{\hat{t}\hat{u}}
&=	\mathcal{G}^{\hat{u}\hat{t}}=
  0,
\qquad
  	\mathcal{G}^{\hat{x}\hat{u}}
  	=\mathcal{G}^{\hat{u}\hat{x}}
=O(u-u_{*}),
\qquad
  	\mathcal{G}^{\hat{t}\hat{x}}
  	=\mathcal{G}^{\hat{x}\hat{t}}
  	=O(1),
\label{eq:invdiagmetric}
}
and the inverse of the metric is not diagonal even at $u=u_{*}$.

However, we will see that only $\mathcal{G}^{\hat{t} \hat{t}}$ and $\mathcal{G}^{\hat{u} \hat{u}}$ contribute to the equations of motion for the fluctuations in the vicinity of $u=u_{*}$, as is given in (\ref{eq:EOMofpsitilde}).
Therefore, the dynamics of the fluctuations in the vicinity of $u=u_{*}$ is governed by $\mathcal{G}^{\hat{t} \hat{t}}$ and $\mathcal{G}^{\hat{u} \hat{u}}$, and the fluctuations observe an effective horizon at $u=u_{*}$ where $\mathcal{G}_{\hat{t} \hat{t}}$ and $\mathcal{G}^{\hat{u} \hat{u}}$ vanish.
In this sense, we regard that the open-string metric provides an analog black hole for the dynamics of the fluctuations whose horizon is located at $u=u_{*}$.

On the new coordinates, derivatives are given by 
\eq{
	\partial_{{\hat{t}}}
&=
	\partial_{t}
	-
	Q^{\hat{x}}_{~t}
	\partial_{x}
,
\qquad
	\partial_{\hat{x}}
=
	\partial_{x}
,
\qquad
	\partial_{\hat{u}}
=
	-
	Q^{\hat{t}}_{~u}
	\partial_{t}
	+
	\roundLR{Q^{\hat{t}}_{~u} Q^{\hat{x}}_{~t}-Q^{\hat{x}}_{~u}}
	\partial_{x}
	+
	\partial_{u}
.
\label{eq:derivatives}
}
The first equation indicates that $-Q^{\hat x}_{~t}$ is understood as the velocity of the analog black hole \cite{Kim:2011qh} from the viewpoint of the original coordinates.
By using the explicit representations of ${G}_{xt}$ and ${G}_{xx}$,
\eq{
	{G}_{xt} 
&=
	-
	\rho
	{J^x}
	\varTheta
,
\qquad
	{G}_{xx} 
=
	\left(
	(2\pi \alpha')^2
	{h}_{xx}^3  	
	V^2
	+
	{\rho^2}
	\right)
	\varTheta
,
\\
	\varTheta
&=
	\frac{
	{ h_{xx}}
	 \absLR{ h_{tt}}
	 -
	(2\pi \alpha')^2
	E_x^2
	}{
	(2\pi \alpha')^2
	{h}_{xx}^3 
	 \absLR{ h_{tt}}
	V^2
	-
	{{J^x}^2} { h_{xx}}
	+
	{\rho^2}
	 \absLR{ h_{tt}}
	}
,
}
we find
\eq{
	-
	Q^{\hat{x}}_{~t}
	(u_*)
=
	-
	\frac{{G}_{xt}}{{G}_{xx}}
	\Bigg|_{u=u_*}
=
	\frac{ \rho J^x }{
	(2\pi \alpha')^2 
	h_{xx}^{3}
	V^2
	+
	{\rho^2}
	}
	\Bigg|_{u=u_*}
=
	{v_m^x}
,
\label{eq:momentumchemicalpotential}
}
where we have used (\ref{eq:vmandrhoj}) in the last equality.
Thus the velocity of the analog black hole, in the sense of mentioned above, is identified with the mean velocity of the charge carriers in the system. We propose (\ref{eq:momentumchemicalpotential}) as a holographic dictionary to give the mean velocity of the charge carriers. 

In Ref.~\cite{Kim:2011qh}, the ``net velocity'' $J_{n_{q}}/\rho$ is defined by using $J_{n_{q}}$ that is the contribution to the current due to the existence of the doped charge carriers dragged by the external electric field. It has been shown that the net velocity agrees with the velocity of the analog black hole at the limit where the pair creation process is negligible (for example, at the large mass limit or at the high charge density limit). 
However, our mean velocity is different from the net velocity. We stress that the mean velocity agrees with the velocity of the analog black hole for arbitrary mass and charge densities. 


\subsection{Spectrum of fluctuations}

Effective temperatures of NESS have been read by analyzing the Hawking radiations from the analog black holes on the probe D-branes~\cite{Sonner:2012if,Nakamura:2013yqa,Hoshino:2014nfa}.
Let us consider Hawking radiations in our analog black hole. We will go along with the analysis given in \cite{Srinivasan:1998ty} where the spectrum of the radiation is obtained by considering a tunneling effect in the semi-classical approximation.

We consider a fluctuation, which we write $\psi$, of a scalar field whose kinetic term is governed by the open string metric\footnote{One such example is the scalar field $\psi$ in Ref.~\cite{Karch:2007pd} that describes the fluctuations of the D7-brane along the compact $S^{2}$ directions.}. Equation of motion for $\psi$
is given by
\eq{
	\partial_a
	\roundLR{
	\sqrt{-\mathcal{G}}
	\mathcal{G}^{ab}
	\partial_b
	\psi
	}
+({\rm other~terms})
=
	0
,
\label{eq:eomforpsitilde}
}
where (other terms) may contain mass term and interactions.
Let us consider the equation of motion in the vicinity of $\hat{u}=\hat{u}_{*}$ (that corresponds to $u=u_{*}$) by making expansion with respect to $(\hat{u}-\hat{u}_*)$.   The open-string metric behaves
$\mathcal{G}^{\hat{t} \hat{t}}\sim -1/[a(\hat{u}-\hat{u}_*)]$ and
$\mathcal{G}^{\hat{u} \hat{u}}\sim (\hat{u}-\hat{u}_*)/b$,
where $a$ and $b$ are constants. Multiplying $(\hat{u}-\hat{u}_*)$ to (\ref{eq:eomforpsitilde}) so that the contribution from $\mathcal{G}^{\hat{t} \hat{t}}$ becomes $O(1)$, we obtain
\eq{
	\squareLR{
	-
	\partial_{\hat{t}}^2
	+
	\frac{a}{b}
	(\hat{u}-\hat{u}_*)
	\partial_{ \hat{u}}
	\roundLR{
	(\hat{u}-\hat{u}_*)
	\partial_{ \hat{u}}
	}
	}
	\psi
=
	0
.
\label{eq:EOMofpsitilde}
}
Here, other contributions in the equation of motion such as the contributions of the off-diagonal components of the open-string metric, as well as those from the mass term and the interactions, are higher-order contributions in the $(\hat{u}-\hat{u}_*)$ expansion.\footnote{The factor $(\hat{u}-\hat{u}_*)$ in the second term are absorbed into $\partial_{ \hat{u}}$ if we employ the tortoise coordinate. We consider only the interactions that are regular at $u=u_*$.}

A solution to (\ref{eq:EOMofpsitilde}) within the WKB approximation is given by
\eq{
	\psi
&=
	\exp
	\squareLR{
	-i \omega' 
	\roundLR{
	\hat t 
	\pm 
	\sqrt{\frac{b}{a}}
	\int 
	\frac{d\hat{u}}{\hat{u}-\hat{u}_*}
	}
	}
,
}
where $\omega'$ is the angular frequency for the fluctuation on the $(\hat{t}, \hat{u})$ coordinates.
The computation of the ratio between the probability of absorption $P_{\mbox{\scriptsize in}}$ and that of emission $P_{\mbox{\scriptsize out}}$ of the fluctuation at $\hat{u}=\hat{u}_*$ is completely parallel to that for the conventional black holes, and we obtain
\eq{
	\frac{
	P_{\mbox{\scriptsize out}}
	}{
	P_{\mbox{\scriptsize in}}
	}
&=
	\exp \roundLR{ -\frac{\omega'}{T_*}  }	
.
\label{eq:probdensities}
}
The effective temperature is given by 
$
	T_{*}
=
	\frac{1}{4\pi}
	\sqrt{\frac{a}{b}}
$ as discussed in \cite{Sonner:2012if,Nakamura:2013yqa,Hoshino:2014nfa}. We use (\ref{eq:derivatives}) to rewrite $\omega'$:
\eq{
	\omega'
&=
	\omega
	+
	Q^{\hat{x}}_{~t}
	k_x
=
	\omega-v_m^x k_x
,
\label{eq:relabetomegaprimeandomega}
}
where $\omega$ is the angular frequency naturally defined on the boundary with respect to the time coordinate $t$, and $k_x$ is the wavenumber of the fluctuation along the $x$ direction on the boundary. Now the spectrum of the fluctuation obeys the distribution function proportional to
\begin{eqnarray}
\exp \roundLR{ -\frac{\omega-v_m^x k_x}{T_*}}.
\label{spectrum} 
\end{eqnarray}
The spectrum is characterized by the mean velocity as well as the effective temperature.

\section{Conclusions and discussion}
\label{sec:discuss}

In the present paper, we have proposed new dictionaries for holographic conductors in which we can compute the carrier densities and the mean velocities of the chare carriers. In the gravity dual, these quantities are not conjugate to the boundary values of the bulk fields, and we cannot apply the conventional GKP-Witten prescription for their estimation. In the present work, we have constructed a phenomenological model of charge transport in which the carrier density and the mean velocity are introduced. 
{Matching the results from the model to those from holography, we have identified the expressions for the carrier density and the mean velocity in a wide range of holographic conductors.}

{In the phenomenological model, we {have} distinguished the friction coefficient for the positive charge carriers ($\gamma_{+}$) and that for the negative charge carriers ($\gamma_{-}$), so that the model can describe effects coming from $\rho$, the difference of the number densities between the positive and the negative charge carriers. 
}
However, we {have} found that $\gamma_{+}=\gamma_{-}$ in general in our holographic models. Furthermore, the friction coefficient {has been} found to be independent of $V$ that reflects the configuration of the probe D-brane.
This means that the friction of the charge carriers in our holographic conductor comes only from the interactions between the charge carrier and the heat bath, and the interactions among the charge carriers do not contribute to the friction within our setup. We understand that this is owing to the probe approximation.

{When we employ the D$p$-brane geometry, we find a characteristic feature that $\Gamma=\gamma\sqrt{1-v^{2}}$, the friction coefficient divided by the Lorentz factor, is a constant in $v$ (hence independent of the external field). Therefore, all the nonlinearity of $\gamma$ is coming from the Lorentz factor in this case.}

This suggests that the highly nonlinear phenomena in charge transport, such as NDC ($\partial J^{x}/\partial E_x < 0$), are owing to the nontrivial behavior of the carrier density.
For $\vec B=0$ cases, for example, (\ref{eq:Jbyv}) gives $J^{x}=n v^{x}$. Then NDC is realized if
\begin{eqnarray}
	\frac{\partial }{\partial E_x}
	\ln n
<
	-
	\frac{\partial }{\partial E_x}
	\ln  {v^x}
.
\label{NDCcondition}
\end{eqnarray}
We know $\frac{\partial }{\partial E_x}\ln {v^x} \geq 0$ since ${v^x}$ is a monotonically increasing function of $E_x$. (\ref{NDCcondition}) shows that NDC is realized if the decrease of $n$ is rapid enough when we increase $E_{x}$.

The carrier density is not a conserved quantity and it should be determined dynamically in the balance between the pair creation and the annihilation of the charge carriers. In general, the probabilities of pair creation and annihilation depend on the mass of the carriers (or the band gap in the context of solid state physics). In the gravity dual, the mass of the carriers gives a boundary condition for the D-brane configuration that determines $V(u)$. Our dictionary (\ref{eq:dtilgene2}) gives the relationship between the carrier density and $V(u_{*})$: the dictionary opens a chance for further investigation of the relationship among the carrier densities, the pair creation/annihilation process of the carriers, and the mass/band gap under the presence of external fields, which may provide us useful information {on the mechanism} of the nonlinear conductivities such as NDC. 
{We {have} provided the dictionary for the mean velocity in terms of the physics of the analog black hole on the worldvolume of the probe D-brane.
We have shown that the mean velocity shifts the spectrum of the fluctuations, in the analysis of the Hawking radiation from the analog black hole. It has already been known that the Hawking temperature of the analog black hole gives the effective temperature of NESS. This means that the analog black holes on the probe D-branes capture fundamental characteristics of the dual NESS. Further studies on the analog black holes in this context will be important for better understanding of NESS. }

\section*{Acknowledgments}
We thank Y. Fukazawa, T. Hayata and S. Kinoshita for fruitful discussions and comments.
The work of S. N. was supported in part by JSPS KAKENHI Grant Number JP16H00810, and
the Chuo University Personal Research Grant. 

\appendix
\renewcommand{\thesection}{\Alph{section}}
\renewcommand{\thesubsection}{\Alph{section}.\arabic{subsection}}
\renewcommand{\theequation}{\Alph{section}.\arabic{equation}}
\section{Derivation of (\ref{pre_sym_condition}), (\ref{eq:gammaunderthecondition1}) and (\ref{eq:dtilapp1})}
\label{sec:derivationofgpeqgm}
We show the derivation of (\ref{pre_sym_condition}), (\ref{eq:gammaunderthecondition1}), and (\ref{eq:dtilapp1}).
We begin with (\ref{eq:vpm}):
\eq{
	\vec v_\pm
&=
	\pm
	\frac{
	(
	\vec E
	\cdot
	\vec B
	)
	\vec B
	+
	\gamma_\pm
	\roundLR{
	\bar \gamma_\pm
	\vec E
	\pm
	\roundLR{
	\vec E
	\times
	\vec B
	}
	}
	}{
	\gamma_\pm
	\roundLR{
	B^2
	+
	\gamma_\pm^2
	}
	}
.
}
Substituting this to the definition of $\vec J$ (\ref{J-def}), we obtain
\eq{
	\vec J
&=
	\rho_+
	\frac{
	\roundN{ \vec{E} \cdot \vec{B} }
	\vec B 
	+
	{\gamma}_+^2
	\vec E 
	+
	{\gamma}_+ 
	\vec{E}\times \vec{B}
	}{
	{\gamma}_+ \left(B^2+{\gamma}_+^2\right)
	}
	+
	\rho_-
	\frac{
	\roundN{ \vec{E} \cdot \vec{B} }
	\vec B
	+
	{\gamma}_-^2
	\vec E 
	-
	{\gamma}_- 
	\vec{E}\times \vec{B} 
	}{
	{\gamma}_- 
	\left(B^2+{\gamma}_-^2\right)
	}
.
\label{eq:curentingeneral}
}
Then we have the following inner products:
\eqna{
	\vec J
	\cdot
	\roundN{	\vec{E}\times \vec{B} }
&=&
	\squareLR{
	\frac{
	\rho_+
	}{
	\left(B^2+{\gamma}_+^2\right)
	}
	-
	\frac{
	\rho_-
	}{
	\left(B^2+{\gamma}_-^2\right)
	}
	}
	\roundN{	\vec{E}\times \vec{B} }^2
,
\label{eq:JdotEcrB}
\\
	\vec{J}\cdot \vec{B}
&=&
	\squareLR{
	\frac{
	\rho_+
	}{
	{\gamma}_+
	}
	+
	\frac{
	\rho_-
	}{
	{\gamma}_- 
	}
	}
	\roundN{ \vec{E} \cdot \vec{B} }
.
\label{eq:JdotB_2}
}
By using (\ref{eq:curentingeneral}), the vector product $\vec J\times \vec{E}$ is given by
\eq{
	\vec{J}\times \vec{E}
&=
	-
	\rho_+
	\frac{
	\roundN{ \vec{E} \cdot \vec{B} }
	\roundN{ \vec{E}\times \vec{B} }	
	-
	{\gamma}_+ 
	\roundN{ \vec{E}\times \vec{B} }	
	\times \vec{E}
	}{
	{\gamma}_+ \left(B^2+{\gamma}_+^2\right)
	}
	-
	\rho_-
	\frac{
	\roundN{ \vec{E} \cdot \vec{B} }
	\roundN{ \vec{E}\times \vec{B} }	
	+
	{\gamma}_- 
	\roundN{ \vec{E}\times \vec{B} }	
	\times \vec{E}
	}{
	{\gamma}_- 
	\left(B^2+{\gamma}_-^2\right)
	}
,
}
and hence we reach the following relationship
\eq{
	\roundN{ \vec{J}\times \vec{E} }
	\cdot
	\roundN{ \vec{E}\times \vec{B} }
&=
	-
	\squareLR{
	\frac{
	\rho_+
	}{
	{\gamma}_+ 
	\left(B^2+{\gamma}_+^2\right)
	}
	+
	\frac{
	\rho_-
	}{
	{\gamma}_- 
	\left(B^2+{\gamma}_-^2\right)
	}
	}
	\roundN{ \vec{E} \cdot \vec{B} }
	\roundN{ \vec E \times \vec{B} }^2
.
\label{eq:JcrE}
}
(\ref{eq:JdotEcrB}), (\ref{eq:JdotB_2}), and (\ref{eq:JcrE}) lead us to (\ref{pre_sym_condition}):
\eq{
&
	\roundN{ \vec{J} \cdot \vec{B} }
	\vec J
	\cdot
	\roundN{ \vec{E}\times \vec{B} }	
	+
	D
	\roundN{ \vec{J}\times \vec{E} }
	\cdot
	\roundN{ \vec{E}\times \vec{B} }	
\\
&=
	\squareLR{
	\frac{
	1
	}{
	{\gamma}_- 
	}
	+
	\frac{
	1
	}{
	{\gamma}_+ 
	}
	}
	\roundLR{
	\frac{
	1
	}{
	\left(B^2+{\gamma}_+^2\right)
	}
	-
	\frac{
	1
	}{
	\left(B^2+{\gamma}_-^2\right)
	}
	}
	\rho_+
	\rho_-
	\roundN{ \vec{E} \cdot \vec{B} }
	\roundN{ \vec{E}\times \vec{B} }^2
.
\label{pre_sym_condition_app}
}

Let us consider the cases where (\ref{pre_sym_condition_app}) vanishes:
\begin{eqnarray}
	\roundN{ \vec{J} \cdot \vec{B} }
	\vec J
	\cdot
	\roundN{ \vec{E}\times \vec{B} }	
	+
	D
	\roundN{ \vec{J}\times \vec{E} }
	\cdot
	\roundN{ \vec{E}\times \vec{B} }
	=0.
\label{sym_condition_ap}	
\end{eqnarray}
We set $\roundN{ \vec{E} \cdot \vec{B} } \roundN{ \vec{E}\times \vec{B} }^2 \not = 0$.
Hence (\ref{sym_condition_ap}) means $\gamma_+=\gamma_- $ or $\rho_+ \rho_- = 0$.
For the cases $\rho_+ \rho_- \neq 0$, we have $\gamma_+=\gamma_- \equiv \gamma$. Then (\ref{eq:JdotEcrB}) and (\ref{eq:JdotB_2}) yield (\ref{eq:gammaunderthecondition1}) and (\ref{eq:dtilapp1}),
\eqna{
	{\gamma}^2
&=&
	\rho
	\frac{
	\roundN{	\vec{E}\times \vec{B} }^2
	}{
	\vec J
	\cdot
	\roundN{	\vec{E}\times \vec{B} }
	}
	-
	B^2
,
\label{eq:gammaunderthecondition}
\\
	n
&=&
	\frac{
	{\gamma}
	\vec{J}\cdot \vec{B}
	}{
	{ \vec{E} \cdot \vec{B} }
	}
,
\label{eq:dtilapp}
}
respectively.
%
The second equality in (\ref{eq:dtilapp1}) is confirmed as follows.
By using (\ref{balance}), we have 
\eq{
	{\gamma} 
	v_{\pm}^2
&=
	\pm
	\vec E \cdot \vec v_\pm
.
\label{eq:relabetgammaandv}
}
Then $\vec J \cdot \vec E $ is found to be
\eq{
	\vec J
	\cdot 
	\vec E
&=
	\rho_+
	{\gamma}_+ 
	v_+^2
	+
	\rho_-
	{\gamma}_-
	v_-^2
,
\label{eq:JdotE}
}
by using (\ref{eq:relabetgammaandv}) and (\ref{J-def}).
Hence, under the condition $\gamma_+=\gamma_-$, we obtain
\eq{
	n
&=
	\frac{
	\vec J
	\cdot 
	\vec E
	}{
    {\gamma}
	v^2
	}
.
\label{eq:Dtilde2}
}

When $\rho_{+}\rho_{-} = 0$, there are three cases: $\rho_{+} = \rho_{-} = 0$; $\rho_{-}\neq 0$ and $\rho_{+}=0$; $\rho_{+}\neq 0$ and $\rho_{-}=0$. When $\rho_{+} = \rho_{-} = 0$, we have no carriers and no chance to define $\gamma_{+}$ and $\gamma_{-}$. We do not consider this trivial case. For the second (third) case, we can define only $\gamma_{+}$ ($\gamma_{-}$), which we write $\gamma$, 
and then (\ref{eq:gammaunderthecondition}), (\ref{eq:dtilapp}), (\ref{eq:Dtilde2}) hold.




\end{document}